\documentclass[preprint,english,letterpaper,showpacs]{revtex4}
\usepackage{babel,amsmath,amssymb}

\begin{document}

\title{Time-Space Noncommutativity: Quantised Evolutions}

\author{A. P. Balachandran}
\email{bal@phy.syr.edu}
\affiliation{Department of Physics, Syracuse University \\
Syracuse, NY, 13244-1130, USA}

\author{T. R. Govindarajan}
\email{trg@imsc.res.in}
\affiliation{The Institute of Mathematical Sciences \\
 C. I. T. Campus Tharamani, Chennai 600 113, India}

\author{A. G. Martins}
\email{amartins@fma.if.usp.br}
\author{P. Teotonio-Sobrinho}
\email{teotonio@fma.if.usp.br}
\affiliation{Instituto de F\'{\i}sica, Universidade de S\~{a}o Paulo \\
C.P. 66318, S\~{a}o Paulo, SP, 05315-970, Brazil}


\preprint{IMSc/04/09/34}
\preprint{SU-4252-801}

\begin{abstract}
In previous work \cite{Bal_Gov_Mol_Paulo}, we developed quantum physics on the Moyal plane with
time-space noncommutativity, basing ourselves on the work of Doplicher
et al. \cite{Doplicher}. Here we extend it to certain noncommutative versions
of the cylinder, $\mathbb{R}^{3}$ and $\mathbb{R}\times S^{3}$. In all these models, only
discrete time translations are possible, a result known before in the
first two cases \cite{Chaichian}-\cite{J_Matschull}. One
striking consequence of quantised time translations is that even
though a time independent Hamiltonian is an observable, in scattering processes, it is
conserved only modulo $\frac{2\pi}{\theta}$, where $\theta$ is the
noncommutative parameter. (In contrast, on a one-dimensional periodic
lattice of lattice spacing $a$ and length $L=Na$, only momentum mod
$\frac{2\pi}{L}$ is observable (and can be conserved).) Suggestions for
further study of this effect are made. Scattering theory is formulated
and an approach to quantum field theory is outlined.
\end{abstract}

\maketitle

\section*{I. Introduction}

Let $x=(x_{0}\,,\,\vec{x})\in\mathbb{R}^{d}$ be coordinates of
$\mathbb{R}^{d}$ with $x_{0}$ and $x_{i}\,(i=1,2,3)$ being its time and spatial
components. The coordinate functions $\hat{x}_{\mu}$ are then defined
by evaluation map:

\begin{equation}\label{evaluation_map}
\hat{x}_{\mu}\,(x)=x_{\mu}\,.
\end{equation}
The algebra generated by $\hat{x}_{\mu}$ with $*$-operation 
$\hat{x}_{\mu}^{*}=\hat{x}_{\mu}$ is the $C^{*}$-algebra
$\mathcal{C}^{0}\,(\mathbb{R}^{d}):=\mathcal{A}_{0}(\mathbb{R}^{d})$
of $\mathbb{R}^{d}$.

The Moyal plane $\mathcal{A}_{\theta}(\mathbb{R}^{d})$ is a
deformation of this algebra where $\hat{x}_{\mu}$ do not commute:

$$
\left[\,\hat{x}_{\mu}\,,\,\hat{x}_{\nu}\,\right]=i\,\theta_{\mu\nu}\mathbb{I}\,,
$$

\begin{equation}\label{Moyal_plane}
\theta_{\mu\nu}=\mbox{real constants},\quad
\theta_{\mu\nu}=-\theta_{\nu\mu}\,.
\end{equation}

In previous work \cite{Bal_Gov_Mol_Paulo}, we developed quantum
mechanics on $\mathcal{A}_{\theta}(\mathbb{R}^{d})$ with time-space
noncommutativity,

\begin{equation}\label{time_space}
\theta_{0i}\neq 0\,,
\end{equation}
assuming for simplicity that spatial coordinates commute,

\begin{equation}\label{commutator_spatial}
\theta_{ij}=0,\quad i,j\,\in\{1,2,3\}
\end{equation}
and basing ourselves on the approach of Doplicher et al.
\cite{Doplicher}. Theory of classical waves and particles on such
spacetimes was also formulated and applied to interference phenomena
\cite{Bal_Gupta}.

In this paper, we continue this line of investigation and study the
following three algebras and their physics.

\subsection*{\it 1) The Noncommutative Cylinder 
$\mathcal{A}_{\theta}\,\left(\mathbb{R}\times S^{1}\right)$}

It is generated by $\hat{x}_{0}$ and $e^{-i\hat{x}_{1}}$ with the
relation

\begin{equation}\label{cylinder_commutator}
\left[\,\hat{x}_{0}\,,\,e^{-i\hat{x}_{1}}\right]=\theta\,e^{-i\hat{x}_{1}}\,.
\end{equation}

\subsection*{\it 2) Noncommutative $\mathbb{R}^{3}$}

The algebra $\hat{e}_{2}$ in this case is the enveloping algebra of
$e_{2}$, the Lie algebra of the Euclidean group $E_{2}$. Spacetime
coordinates $\hat{x}_{\mu}$ form a basis of $e_{2}$ and fulfill the
commutation relations

$$
\left[\,\hat{x}_{i}\,,\,\hat{x}_{j}\right]=0\,,\quad
\left[\,\hat{x}_{0}\,,\,\hat{x}_{i}\right]=i\theta\,\epsilon_{ij}\hat{x}_{j}\,,
$$

\begin{equation}\label{gravity_commutator}
\epsilon_{ij}=-\epsilon_{ji}\,,\quad
\epsilon_{12}=1\,.
\end{equation}
Thus $\hat{x}_{i}$ are identified with translations and
$\hat{x}_{0}/\theta$ is the canonically normalised angular
momentum $J$:

\begin{equation}\label{angular_momentum}
e^{i2\pi J}\hat{x}_{\mu}e^{-i2\pi J}=\hat{x}_{\mu}\,.
\end{equation}

The Lie algebra $e_{2}$ is a contraction of $so(2,1)$, the Lie algebra
of $SO(2,1)$. The latter and its enveloping algebra have occurred as
spacetime algebras in 2+1 gravity \cite{Matschull}. (See also
\cite{Hooft}-\cite{J_Matschull} and also \cite{Bal_Chandar} in this connection.)

\subsection*{\it 3) The Noncommutative $\mathbb{R}\times S^{3}$, 
$\mathcal{A}_{\theta}\,\left(\mathbb{R}\times S^{3}\right)$}

We can represent $S^{3}=\left\langle
x\in\mathbb{R}^{4}\,:\,\sum_{\lambda}x^{2}_{\lambda}=1\right\rangle$ by 
$SU(2)$ matrices:

\begin{equation}
x_{0}\mathbb{I}+i\vec{\tau}\cdot\vec{x}\,\,\in SU(2)\,,
\end{equation}
where $\mathbb{I}$ is the $2\times 2$ unit matrix and $\tau_{i}$ are
Pauli matrices.
In this way we identify $S^{3}$ and $SU(2)$. Left- and right- regular
representations of $SU(2)$ act on functions
$\mathcal{C}^{\infty}\,\left(SU(2)\right)$ on $SU(2)$.

Let $su(2)$ be the Lie algebra of $SU(2)$ with conventional angular
momentum operators $J_{i}$. Then in particular, $J_{3}$ has a right
action $J_{3}^{R}$ on $\mathcal{C}^{\infty}\,\left(SU(2)\right)$:

\begin{equation}\label{su_2}
\left(e^{i\theta
J_{3}^{R}}\,\hat{f}\right)\,(g)=\hat{f}\,\left(g\,e^{i\theta
J_{3}/2}\right)
\end{equation}
for $\hat{f}\in\mathcal{C}^{\infty}\,\left(SU(2)\right)$ and $g\in SU(2)$.

In the algebra
$\mathcal{A}_{\theta}\,\left(\mathbb{R}\times S^{3}\right)$, the
spatial slice $S^{3}$ is represented by the commutative algebra 
$\mathcal{C}^{\infty}\,\left(SU(2)\right)$, and time $\hat{x}_{0}$ is
identified with $2\theta J_{3}^{R}$ in the following way:

\begin{equation}\label{non_s3}
\left(e^{i\omega\hat{x}_{0}}\,\hat{f}\,e^{-i\omega\hat{x}_{0}}\right)(s):=
\left(e^{i\omega 2\theta J_{3}^{R}}\,\hat{f}\right)(s)\,.
\end{equation}

Cases 1) and 3) are actually very similar. In case 1), the spatial slice
has algebra $C^{\infty}\left(S^{1}\right)$ and
$J=\hat{x}_{0}/\theta$ is the canonically normalised generator
of rotations: $e^{i2\pi J}\hat{\alpha}\,e^{-i2\pi J}=\hat{\alpha}$, for 
$\hat{\alpha}\in\mathcal{A}_{\theta}\left(\mathbb{R}\times S^{1}\right)$.

In all these cases, time translations get quantised in units of
$\theta$ in quantum physics. This result is known for cases 1) and
2) \cite{Chaichian}-\cite{J_Matschull}. It comes from the
fact that the spectrum $\mbox{spec}\,J$ or $\mbox{spec}\,J_{3}$ of $J$ or $J_{3}$ in an irreducible
representation of the associated algebra is spaced in units of $\theta$. We
will prove it fully as we go along.

Using a different approach, a model with quantised evolution was also
constructed in \cite{Bal_Chandar}.

There are generalisations of these constructions to manifolds
$\mathbb{R}\times M$ where $\mathbb{R}$ accounts for time and
$M$ is the spatial slice, provided $M$ admits a $U(1)$ action. If $J$ is its
generator on $\mathcal{C}^{\infty}(M)$, we can set
$\hat{x}_{0}=\theta J$ and get an algebra
$\mathcal{A}_{\theta}\left(\mathbb{R}\times M\right)$ with quantised
evolution.

The mathematical approach to noncommutativity in this paper is similar
to that of Rieffel, Connes, Landi and others \cite{Rieffel}-\cite{Connes_Landi}. We have
drawn much inspiration from their work.

After reviewing \cite{Bal_Gov_Mol_Paulo} in the next section, we will
study the three preceding examples in the subsequent sections. Issues
related to energy nonconservation and also scattering and quantum
field theory are taken up after that.

\section*{II. The Noncommutative Plane}

\subsection*{\it 1. Generalities}

The noncommutative or Moyal plane
$A_{\theta}\left(\mathbb{R}^{d}\right)$ is based on the commutation
relation (\ref{Moyal_plane}). We outline how to do quantum physics on
this algebra, summarising \cite{Bal_Gov_Mol_Paulo}. It is enough to
consider $d=2$. Then we can write, without loss of generality,

$$
\left[\hat{x}_{\mu}\,,\,\hat{x}_{\nu}\right]=i\theta\epsilon_{\mu\nu},\quad
\mu,\nu\in\,\{0\,,\,1\},
$$

\begin{equation}\label{nc_gravity}
\epsilon_{\mu\nu}=-\epsilon_{\nu\mu}\,,\quad
\epsilon_{01}=-\epsilon_{10}=1\,,\quad \theta\geq 0\,.
\end{equation}

Since $\hat{x}_{1}$ and $-\hat{x}_{0}/\theta$ have the
commutation relation of position and momentum, we can realise this
algebra irreducibly on $L^{2}\left(\mathbb{R}\right)$ as in elementary
quantum mechanics.

For quantum mechanics on
$\mathcal{A}_{\theta}\left(\mathbb{R}^{2}\right)$, elements of 
$\mathcal{A}_{\theta}\left(\mathbb{R}^{2}\right)$ itself constitute
the ``wave functions'' or ``vector states''. Then for 
$\hat{\alpha}\in\mathcal{A}_{\theta}\left(\mathbb{R}^{2}\right)$, we
have two linear operators $\hat{\alpha}^{L,R}$ on these vector states:

\begin{equation}
\hat{\alpha}^{L}\hat{\psi}=\hat{\alpha}\hat{\psi}\,,\quad 
\hat{\alpha}^{R}\hat{\psi}=\hat{\psi}\hat{\alpha}\,,\quad
\hat{\psi}\in \mathcal{A}_{\theta}(\mathbb{R}^{2})\,.
\end{equation}
Their difference $ad\,\hat{\alpha}$ gives the adjoint action of
$\hat{\alpha}$:

\begin{equation}
ad\,\hat{\alpha}=\hat{\alpha}^{L}-\hat{\alpha}^{R},\quad
ad\,\hat{\alpha}\,\hat{\psi}=[\hat{\alpha}\,,\,\hat{\psi}]\,.
\end{equation}

For $\theta=0$, we have time- and space- translation generators
$i\partial_{t}$ , $-i\frac{\partial}{\partial x_{1}}$. The latter is
the momentum. Their analogues here are $\hat{P}_{0}$ , 
$\hat{P}_{1}$ where

\begin{eqnarray}
\hat{P}_{0}=-\frac{1}{\theta}ad\,\hat{x}_{1}\,,\quad\hat{P}_{1}=
-\frac{1}{\theta}ad\,\hat{x}_{0}\,,
\end{eqnarray}

\begin{eqnarray}
\hat{P}_{\mu}\hat{x}_{\nu}=i\eta_{\mu\nu}\mathbb{I}\,,\quad
\eta_{\mu\nu}=0\quad\mbox{if}\quad\mu\neq\nu\,,\quad\eta_{00}=-\eta_{11}=1\,.
\end{eqnarray}

\subsection*{\it 2. The Inner Product}

The next step is to find an inner product on the vector states. There
are several (equivalent) possibilities \cite{Bal_Gov_Mol_Paulo}. We
describe one here.

Let 

\begin{equation}\label{element_tilde}
\hat{\psi}=\int d^{2}k\,\tilde{\psi}(k)\,e^{ik_{1}\hat{x}_{1}}
\,e^{ik_{0}\hat{x}_{0}}\,.
\end{equation}
We define its symbol $\psi$, which is a complex function on
$\mathbb{R}^{2}=\{x=(x_{0}\,,\,x_{1})\}$, by

\begin{equation}\label{define_symbol}
\psi=\int d^{2}k\,\tilde{\psi}(k)\,e^{ik_{1}x_{1}}
\,e^{ik_{0}x_{0}}\,.
\end{equation}
The inner product $(\hat{\psi}\,,\,\hat{\eta})_{x_{0}}$ of
two vector states $\hat{\psi}$, $\hat{\eta}$ is then

\begin{equation}\label{inner}
(\hat{\psi}\,,\,\hat{\eta})_{x_{0}}=
\int dx_{1}\,\psi^{*}(x_{0}\,,x_{1})
\,\eta(x_{0}\,,x_{1})\,,
\end{equation}
$\eta$ being the symbol of $\hat{\eta}$. It depends on $x_{0}$.

\subsection*{\it 3. The Hilbert Space}

If $\psi$ vanishes at ``time'' $x_{0}$ and all $x_{1}$, $\hat{\psi}$ is a null
vector in this inner product. The set $\mathcal{N}$ of null vectors in
this inner product is thus large. The inner product depends as well on $x_{0}$. Both
the nontrivial null vectors and dependence on $x_{0}$ can be
eliminated by imposing the Schr\"{o}dinger equation or
``constraint''. The completion of the resultant vector states in the
scalar product is the quantum Hilbert space $\mathcal{H}$.

Let $\hat{H}$ be a ``time-independent'' Hamiltonian hermitian in the
inner product:

\begin{equation}
[\hat{P}_{0}\,,\,\hat{H}]=0 ,\quad 
(\hat{\psi}\,,\,\hat{H}\hat{\eta})_{x_{0}}=
(\hat{H}\hat{\psi}\,,\,\hat{\eta})_{x_{0}}\,.
\end{equation}
Let $\mathcal{C}$ be the vectors fulfilling the Schr\"{o}dinger
equation:

\begin{equation}\label{vectors_fulfilling}
\mathcal{C}=\left\langle\hat{\psi}\in\mathcal{A}_{\theta}\left(\mathbb{R}^{2}\right):
\left(\hat{P}_{0}-\hat{H}\right)\hat{\psi}=0\right\rangle\,.
\end{equation}
Then since

\begin{equation}
\left(e^{-i\hat{P}_{0}\tau}\hat{\psi}\,,\,e^{-i\hat{P}_{0}\tau}\hat{\eta}\right)_{x_{0}}=
\left(e^{-i\hat{H}\tau}\hat{\psi}\,,\,e^{-i\hat{H}\tau}\hat{\eta}\right)_{x_{0}}=
\left(\hat{\psi}\,,\,\hat{\eta}\right)_{x_{0}}
\end{equation}
for $\hat{\psi}\,,\,\hat{\eta}\in\mathcal{C}$, we can see that the
inner product is independent of $x_{0}$ for vectors in
$\mathcal{C}$. So we write $(\hat{\psi}\,,\hat{\eta})_{x_{0}}$ as $(\hat{\psi}\,,\hat{\eta})$.

Also if $\psi_{\tau}$ is the symbol of
$e^{-i\hat{P}_{0}\tau}\hat{\psi}$, then (\ref{element_tilde}),
(\ref{define_symbol}) show that 
$\psi_{\tau}(x_{0}\,,\,x_{1})=\psi(x_{0}+\tau\,,\,x_{1})$.
Hence if $\hat{\psi}\in\mathcal{C}$ and $\psi$ is zero at $x_{0}$ and
all $x_{1}$, then it is identically zero. From (\ref{define_symbol}),
$\tilde{\psi}$ is then zero, and so by (\ref{element_tilde}), $\hat{\psi}=0$. So we
also get

\begin{equation}
\mathcal{N}\cap\mathcal{C}=\{0\}
\end{equation}
as claimed.

There is a simple solution for the Schr\"{o}dinger constraint for
time-independent $\hat{H}$:

\begin{equation}\label{simple_solution}
\hat{\psi}\in \mathcal{C}\Longrightarrow
\hat{\psi}=e^{-i\hat{H}\hat{x}_{0}^{R}}\hat{\chi}\left(\hat{x}_{1}\right)\,.
\end{equation}
The vector $\hat{\chi}\left(\hat{x}_{1}\right)$ has no dependence on
$\hat{x}_{0}$, that is, commutes with $\hat{x}_{1}$.

We can extend the discussion to time-dependent, but still hermitean,
Hamiltonians. For details, see \cite{Bal_Gov_Mol_Paulo}.

\subsection*{\it 4. On Positive Maps}

A positive map $S$ on a $*$-algebra $\mathcal{A}$ is a linear map
$S:\hat{\alpha}\in\mathcal{A}\longmapsto S(\hat{\alpha})\in\mathbb{C}$
with the properties

$$
S(\hat{\alpha}^{*})=S(\hat{\alpha})^{*}\,,
$$

\begin{equation}
S\left(\hat{\alpha}^{*}\hat{\alpha}\right)\geq 0\,.
\end{equation}
Given such a map $S$, we can define an inner product
$\left(\cdot\,,\,\cdot\right)$ on $\mathcal{A}$:
$(\hat{\alpha}\,,\,\hat{\beta})=S(\hat{\alpha}^{*}\hat{\beta})$.

The inner product (\ref{inner}) comes from the following positive map
$S_{x_{0}}$:

\begin{equation}
S_{x_{0}}(\hat{\psi})=\int dx_{1}\,\psi(x_{0}\,,\,x_{1})\,.
\end{equation}

Positive maps, like traces, can substitute for integration on
algebras, and are useful for formulating physical theories on
noncommutative spacetimes.

\section*{III. The Noncommutative Cylinder}

The noncommutative cylinder
$\mathcal{A}_{\theta}\left(\mathbb{R}\times S^{1}\right)$ has been
considered in great detail by Chaichian et al. \cite{Chaichian},
especially as regards its quantum field theory aspects. They have
pointed out and emphasised that time gets quantised on
$\mathcal{A}_{\theta}\left(\mathbb{R}\times S^{1}\right)$
(see also \cite{Bal_Chandar}) and studied
the impact of this quantisation on causality and unitarity. Below, we
review how this quantisation comes about and develop quantum physics on
$\mathcal{A}_{\theta}\left(\mathbb{R}\times S^{1}\right)$. We do not
encounter problems with unitarity. 

For $\theta=0$, there is a close relation between
$C^{\infty}\left(\mathbb{R}\times\mathbb{R}\right)$ and the functions
$C^{\infty}\left(\mathbb{R}\times S^{1}\right)$ on a cylinder. The
former is generated by coordinate functions $\hat{x}_{0}$ and
$\hat{x}_{1}$, and the latter by $\hat{x}_{0}$ and $e^{i\hat{x}_{1}}$,
$e^{i\hat{x}_{1}}$ being invariant under the $2\pi$-shifts
$\hat{x}_{1}\rightarrow\hat{x}_{1}\pm2\pi$. Following this idea, we
can regard the noncommutative $\mathbb{R}\times S^{1}$ algebra
$\mathcal{A}_{\theta}\left(\mathbb{R}\times S^{1}\right)$ as generated
by $\hat{x}_{0}$ and $e^{i\hat{x}_{1}}$ with the defining relation 

$$
e^{i\hat{x}_{1}}\hat{x}_{0} =
\hat{x}_{0}e^{i\hat{x}_{1}}+ \theta e^{i\hat{x}_{1}} \,,
$$
following from (\ref{nc_gravity}).

For $C^{\infty}\left(\mathbb{R}\times S^{1}\right)$, the momentum
$\hat{p}_{1}$ is the differential operator defined by 
\begin{equation}
\left[\hat{p}_{1},e^{i\hat{x}_{1}}\right] =
e^{i\hat{x}_{1}} \,.
\label{p1}
\end{equation}
By evaluating (\ref{p1}) at $x_{1}$, we can write it in the usual
way:  $\left[-i\frac{\partial}{\partial x_{1}},e^{ix_{1}}\right]=e^{ix_{1}}$.

It follows from (\ref{p1}) that 
\begin{equation}\label{center}
e^{i2\pi\hat{p}_{1}}e^{i\hat{x}_{1}}e^{-i2\pi\hat{p}_{1}} =
e^{i\hat{x}_{1}} \,.
\end{equation}
So $e^{i2\pi\hat{p}_{1}}$ is in the center of the algebra generated
by $\hat{p}_{1}$, $e^{i\hat{x}_{1}}$ with the relation (\ref{p1}).
In an irreducible representation (IRR), it is a phase $e^{i\varphi}$
times $\mathbb{I}$.
The spectrum of $\hat{p}_{1}$ in an IRR is hence 
\begin{equation}
\textrm{spec} \, \hat{p}_{1}=\mathbb{Z}+\frac{\varphi}{2\pi}
\equiv \left\{n+\frac{\varphi}{2\pi}:n\in\mathbb{Z}\right\}\,.
\end{equation}
Its domain $\mathcal{D}_{\varphi}(\hat{p}_{1})$ in such an IRR is spanned by quasi-periodic
functions $\chi_{n}$:  
\[
\chi_{n} =
e^{i\left(n+\frac{\varphi}{2\pi}\right)\hat{x}_{1}} \,\, , \quad
n \in \mathbb{Z} \,\, ,
\]
 \begin{equation}
\chi_{n}(\hat{x}_{1}+2\pi)=e^{i\varphi}\chi_{n}(\hat{x}_{1}) \,.
\label{quasi-periodic}
\end{equation}

If for example
\begin{equation}
H = \frac{\hat{p}_{1}^{2}}{2m}
\end{equation}
is the Hamiltonian, its domain $\mathcal{D}_{\varphi}(H)$ fulfilling the
Schr\"{o}dinger constraint as well is spanned by
\begin{equation}
\psi_{n} = \chi_{n}e^{-iE_{n}\hat{x}_{0}} \,,
\end{equation}
with $\psi_{n}$ being eigenstates of $H$:
\begin{equation}
H\psi_{n} = E_{n}\psi_{n} \,\,,
\end{equation}

\begin{equation}
E_{n} =
\frac{1}{2m}\left(n+\frac{\varphi}{2\pi}\right)^{2} \,.
\label{omega_n}
\end{equation}
The quantity $\varphi$ is generally interpreted as the flux through
the circle. 

For the noncommutative cylinder, (\ref{center}) generalises
in a striking manner:
\begin{equation}
e^{-i\frac{2\pi}{\theta}\hat{x}_{0}} e^{i\hat{x}_{1}}
e^{i\frac{2\pi}{\theta}\hat{x}_{0}} =
e^{i\hat{x}_{1}}\,.
\label{nc_cylinder_result}
\end{equation}
Hence in an IRR,
\begin{equation}
e^{-i\frac{2\pi}{\theta}\hat{x}_{0}} =
e^{-i\varphi}\mathbb{I} \,,
\end{equation}
so that for the spectrum $\textrm{spec} \, \hat{x}_{0}$ of
$\hat{x}_{0}$ in an IRR, we have,
\begin{equation}
\textrm{spec} \, \hat{x}_{0} = 
\theta\mathbb{Z}+\frac{\theta\varphi}{2\pi} =
\theta\left(\mathbb{Z}+\frac{\varphi}{2\pi}\right)
\equiv \left\{\theta \left(n+\frac{\varphi}{2\pi}\right):n\in\mathbb{Z}\right\} \,.
\label{spectrum_cylinder}
\end{equation}

We can realise $\mathcal{A}_{\theta}\left(\mathbb{R}\times S^{1}\right)$
irreducibly in the auxiliary Hilbert space $L^{2}\left(S^{1},dx_{1}\right)$.
It has the scalar product given by
\begin{equation}
\left(\alpha,\beta\right) =
\int_{0}^{2\pi}dx_{1} \,
{\alpha}^{*}\left(e^{ix_{1}}\right)\beta \left(e^{ix_{1}}\right)
\,\, , \,\,
\alpha,\beta\in L^{2}\left(S^{1},dx_{1}\right) \,.
\end{equation}
On this space, $e^{i\hat{x}_{1}}$ acts by evaluation map,
\begin{equation}
\left(e^{i\hat{x}_{1}}\alpha\right)\left(e^{ix_{1}}\right) =
e^{ix_{1}}\alpha\left(e^{ix_{1}}\right) \,,
\end{equation}
while $\hat{x}_{0}/\theta$ acts like the $\theta=0$ momentum with
domain $D_{\varphi}(\hat{p}_{1})$.

We denote this particular representation of 
$\mathcal{A}_{\theta}\left(\mathbb{R}\times S^{1}\right)$ as
$\mathcal{A}_{\theta}\left(\mathbb{R}\times S^{1}\,,e^{i\frac{\varphi}{2\pi}}\right)$.

Let us examine
$\mathcal{A}_{\theta}\left(\mathbb{R}\times S^{1}\,,e^{i\frac{\varphi}{2\pi}}\right)$ 
more closely. We can regard it as generated by $e^{i\hat{x}_{1}}$ and
$e^{i\omega\hat{x}_{0}}$ where $\omega$ is real. Now because of the
spectral result (\ref{spectrum_cylinder}),
\begin{equation}
e^{i\left(\omega+\frac{2\pi}{\theta}\right)\hat{x}_{0}} =
e^{i\varphi}e^{i\omega\hat{x}_{0}} \,.
\end{equation}
Thus elements of
$\mathcal{A}_{\theta}\left(\mathbb{R}\times S^{1}\,,
e^{i\frac{\varphi}{2\pi}}\right)$
are quasiperiodic in $\omega$ just as $\chi_{n}$ is quasiperiodic
in $\hat{x}_{1}$, and we can restrict $\omega$ to its fundamental
domain:
\begin{equation}
\omega\in\left[-\frac{\pi}{\theta},\frac{\pi}{\theta}\right]
\,.
\end{equation}
The general element of $\mathcal{A}_{\theta}\left(\mathbb{R}\times
S^{1}\,,e^{i\frac{\varphi}{2\pi}}\right)$ is thus
\begin{equation}
\hat{\alpha} =
\sum_{n\in\mathbb{Z}} \int_{-\frac{\pi}{\theta}}^{+\frac{\pi}{\theta}}
d\omega \,
\tilde{\alpha}_{n}(\omega)e^{in\hat{x}_{1}}
e^{i\omega\hat{x}_{0}}\,,
\end{equation}
as first discussed by Chaichian et al. \cite{Chaichian}.

\subsection*{\it 1. Positive Maps and Inner Products}

A positive map on
$\mathcal{A}_{\theta}\left(\mathbb{R}\times S^{1}\,,
e^{i\frac{\varphi}{2\pi}}\right)$
can be found from symbol calculus. Since the spectrum of
$\hat{x}_{0}$ is $\theta\left(\mathbb{Z}+\frac{\varphi}{2\pi}\right)$
and the spectrum of $e^{i\hat{x}_{1}}$ is $S^{1}$, the symbol of
$\hat{\alpha}$ is a function $\alpha$ on
$\theta\left(\mathbb{Z}+\frac{\varphi}{2\pi}\right)\times S^{1}$:
\begin{equation}
\alpha: \theta\left(\mathbb{Z}+\frac{\varphi}{2\pi}\right) \times
S^{1} \rightarrow \mathbb{C}\,.
\end{equation} 
It is defined by
\begin{equation}
\alpha\left(\theta\left(m+\frac{\varphi}{2\pi}\right),
e^{ix_{1}}\right)=\sum_{n\in\mathbb{Z}}
\int_{-\frac{\pi}{\theta}}^{+\frac{\pi}{\theta}}d\omega\,
\tilde{\alpha}_{n}(\omega)e^{inx_{1}}
e^{i\omega\theta\left(m+\frac{\varphi}{2\pi}\right)}\,.
\end{equation}

Before proceeding, we show that $\hat{\alpha}$ determines $\tilde{\alpha}_{n}$
and hence $\alpha$ uniquely, so that the map
$\hat{\alpha}\rightarrow\alpha$ is well-defined. We show also the
converse, that $\alpha$ determines $\tilde{\alpha}_{n}$ and hence
$\hat{\alpha}$ uniquely, so that the map
$\hat{\alpha}\rightarrow\alpha$ is bijective. 

Let $\left|n\right\rangle $ be the normalised eigenstates of
$\hat{x}_{0}$:
\begin{equation}\label{norm_eigen}
\hat{x}_{0}\left|n\right\rangle = 
\theta\left(n+\frac{\varphi}{2\pi}\right)\left|n\right\rangle
\,\, , \,\,
\left\langle m|n\right\rangle  = \delta_{mn} \,\, , \quad
n\in\mathbb{Z}\,.
\end{equation}
Then
\begin{equation}
e^{i\hat{x}_{1}}\left|n\right\rangle =
\left|n-1\right\rangle \,.
\end{equation}
Therefore
\begin{equation}
\left\langle m\right|\hat{\alpha}\left|n\right\rangle =
\int_{-\frac{\pi}{\theta}}^{+\frac{\pi}{\theta}}d\omega\,
\tilde{\alpha}_{n-m}(\omega)e^{i\omega\theta
\left(n+\frac{\varphi}{2\pi}\right)}\,,
\end{equation}
and since 
\begin{equation}
\frac{\theta}{2\pi}\sum_{n}e^{i(\omega-\omega')\theta n} =
\delta(\omega-\omega') \,,
\end{equation}
we find
\begin{equation}\label{coefficient}
\frac{\theta}{2\pi}\sum_{n}
e^{-i\omega\theta\left(n+\frac{\varphi}{2\pi}\right)}
\left\langle n-m \right|\hat{\alpha}\left|n\right\rangle =
\tilde{\alpha}_{m}(\omega) \,.
\end{equation}

The inverse map follows similarly:
\begin{equation}
\tilde{\alpha}_{n}(\omega) =
\frac{\theta}{(2\pi)^{2}}\sum_{m}
e^{-i\omega\theta\left(m+\frac{\varphi}{2\pi}\right)}\int_{0}^{2\pi}dx_{1}\,
e^{-inx_{1}}\alpha\left(\theta\left(m+\frac{\varphi}{2\pi}\right),
e^{ix_{1}}\right)\,.
\end{equation}

Our positive map is $S_{\theta\left(m+\frac{\varphi}{2\pi}\right)}$:
\begin{equation}
S_{\theta\left(m+\frac{\varphi}{2\pi}\right)}\left(\hat{\alpha}\right)
= \int_{0}^{2\pi}dx_{1} \,
\alpha\left(\theta\left(m+\frac{\varphi}{2\pi}\right),
e^{ix_{1}}\right) \,.
\end{equation}
Just as in (\ref{inner}), we then have, for inner
product,
\begin{eqnarray}\label{module_inner_product}
\left(\hat{\alpha},\hat{\beta}\right)_{\theta\left(m+\frac{\varphi}{2\pi}\right)}&=&
S_{\theta\left(m+\frac{\varphi}{2\pi}\right)}
\left(\hat{\alpha}^{*}\hat{\beta}\right)\nonumber\\
&=&\int_{0}^{2\pi}dx_{1}\,\alpha^{*}
\left(\theta\left(m+\frac{\varphi}{2\pi}\right),
e^{ix_{1}}\right)
\beta\left(\theta\left(m+\frac{\varphi}{2\pi}\right),
e^{ix_{1}}\right)\,.
\end{eqnarray}

There are other possibilities for inner product such as the one based
on coherent states.
The equivalence of theories based on different inner products is
discussed in \cite{Bal_Gov_Mol_Paulo}.

\subsection*{\it 2. Spectrum of Momentum}

We can infer the spectrum of the momentum operator $\hat{P}_{1}$ when
it acts on $\mathcal{A}_{\theta}\left(\mathbb{R}\times
S^1,e^{i\frac{\varphi}{2\pi}}\right)$. Since this algebra allows for only integral powers of
$e^{i\hat{x}_{1}}$, and 

\begin{equation}
\hat{P}_{1}e^{in\hat{x}_{1}}=ne^{in\hat{x}_{1}},
\end{equation} 

\noindent
we have

\begin{equation}
\mbox{spec}\, \hat{P}_{1}=\mathbb{Z}\,.
\end{equation}

\noindent
The flux term is 0 in this spectrum.

For the construction of a Hilbert space, we do not need this algebra.
It is enough to have an $\mathcal{A}_{\theta}\left(\mathbb{R}\times
S^1,e^{i\frac{\varphi}{2\pi}}\right)$-module which can be consistently treated. Such a module is

\begin{equation}
\mathcal{A}_{\theta}\left(\mathbb{R}\times S^{1},e^{i\frac{\varphi}{2\pi}},
e^{i\frac{\psi}{2\pi}}\right)=\left< \hat{\gamma}=e^{i\frac{\psi}{2\pi}\hat{x}_{1}}\sum_{n\in
\mathbb{Z}}\int_{-\frac{\pi}{\theta}}
^{\frac{\pi}{\theta}}d\omega\tilde{\gamma}_{n}(\omega)
e^{in\hat{x}_{1}}e^{i\omega\hat{x}_{0}}\right>\,.
\end{equation}

\noindent
The eigenvalues of $\hat{P}_{1}$ are now shifted by $\frac{\psi}{2\pi}$:

\begin{equation}\label{eigenvalues}
\hat{P}_{1}e^{i\frac{\psi}{2\pi}\hat{x}_{1}}e^{in\hat{x}_{1}}=\left(n+\frac{\psi}{2\pi}\right)
e^{i\frac{\psi}{2\pi}\hat{x}_{1}}e^{in\hat{x}_{1}}\,,\,n\in\mathbb{Z}\,.
\end{equation}

\noindent
So we now have a flux term $\frac{\psi}{2\pi}$.

We have to check that $\mathcal{A}_{\theta}\left(\mathbb{R}\times S^{1},e^{i\frac{\varphi}{2\pi}},
e^{i\frac{\psi}{2\pi}}\right)$ also has an inner product. That is so
because if 

\begin{equation}
\hat{\gamma},\hat{\delta}\in\mathcal{A}_{\theta}\left(\mathbb{R}\times S^{1},e^{i\frac{\varphi}{2\pi}},
e^{i\frac{\psi}{2\pi}}\right)\,, 
\end{equation} 

\noindent
then

\begin{equation}
\hat{\gamma}^{*}\hat{\delta}\in\mathcal{A}_{\theta}\left(\mathbb{R}\times S^{1},e^{i\frac{\varphi}{2\pi}}\right), 
\end{equation} 

\noindent
(the $\psi$-dependent factors $e^{i\frac{\psi}{2\pi}\hat{x}_{1}}$
cancelling out), so that the inner product is still like
(\ref{module_inner_product}):

\begin{equation}\label{same_inner_product}
(\hat{\gamma},\hat{\delta})_{\theta\left(m+\frac{\varphi}{2\pi}\right)}
=S_{\theta\left(m+\frac{\varphi}{2\pi}\right)}
(\hat{\gamma}^{*}\hat{\delta})\,.
\end{equation}

It is interesting that the flux terms in time and momentum can be
different.

We remark that the Schr\"{o}dinger constraint below does not alter the
spectrum of $\hat{P}_{1}$.

\subsection*{\it 3. The Schr\"{o}dinger Constraint}

\subsection*{\it a) The Time-Independent Hamiltonian}

Since
\begin{equation}
i\partial_{x_{0}}e^{i\omega\hat{x}_{0}} =
-\omega e^{i\omega\hat{x}_{0}}
\end{equation}
is not quasiperiodic in $\omega$, continuous time translations and
the Schr\"{o}dinger constraint in the original form cannot be defined
on $\mathcal{A}_{\theta}\left(\mathbb{R}\times S^{1}\right)$.

But translation of $\hat{x}_{0}$ by $\pm\theta$ leaves its spectrum
intact. Hence the operator
\begin{equation}
e^{-i\theta\left(i\partial_{x_{0}}\right)}=
e^{i ad\,\hat{x}_{1}} \,,
\end{equation}
and its integral powers act on
$\mathcal{A}_{\theta}\left(\mathbb{R}\times S^{1}\right)$. The
conventional Schr\"{o}dinger constraint is thus changed to a discrete
Schr\"{o}dinger constraint. In the time-independent case when the
Hamiltonian can be written as
$\hat{H}\left(e^{i\hat{x}_{1}^{L}},\hat{P}_{1}\right)$, the family of
vector states constrained by the discrete Schr\"{o}dinger equation is
\begin{equation}\label{discrete_constraint}
\tilde{\mathcal{H}}_{\theta}\left(e^{i\frac{\varphi}{2\pi}},e^{i\frac{\psi}{2\pi}}\right)=
\left\{ \hat{\psi}\in\mathcal{A}_{\theta}\left(\mathbb{R}\times
S^{1},e^{i\frac{\varphi}{2\pi}},e^{i\frac{\psi}{2\pi}}\right)
:e^{-i\theta\left(i\partial_{x_{0}}\right)}\hat{\psi}=
e^{-i\theta\hat{H}}\hat{\psi}\right\} \,.
\end{equation}
It has solutions 
\begin{equation}\label{solut}
\hat{\psi} =
e^{-i\hat{x}_{0}^{R}\hat{H}\left(e^{i\hat{x}_{1}^{L}},\hat{P}_{1}\right)}
e^{i\frac{\psi}{2\pi}\hat{x}_{1}}\hat{\chi}\left(e^{i\hat{x}_{1}}\right) \,,
\end{equation}
just as in (\ref{simple_solution}).

\subsection*{\it b) The Time-Dependent Hamiltonian}

The time-dependent Hamiltonian is 

\begin{equation}\label{time_dependent_hamiltonian}
\hat{\hat{H}}\left(\hat{x}_{0}^{L},\hat{x}_{0}^{R},e^{i\hat{x}_{1}^{L}},\hat{P}_{1}\right)
\end{equation}

\noindent 
and the Schr\"{o}dinger constraint (\ref{discrete_constraint}) defining
$\tilde{\mathcal{H}}_{\theta}\left(e^{i\frac{\varphi}{2\pi}},e^{i\frac{\psi}{2\pi}}\right)$
is intact. We can solve this constraint as follows.

Write

\begin{equation}
\hat{\hat{H}}\left(\hat{x}_{0}^{L},\hat{x}_{0}^{R},e^{i\hat{x}_{1}^{L}},\hat{P}_{1}\right)=
\hat{\hat{H}}\left(-\theta\hat{P_{1}}+\hat{x}_{0}^{R},\hat{x}_{0}^{R},
e^{i\hat{x}_{1}^{L}},\hat{P}_{1}\right)\equiv
\hat{H}\left(\hat{x}_{0}^{R},e^{i\hat{x}_{1}^{L}},\hat{P}_{1}\right)\,.
\end{equation}
We can try to replace $\hat{x}_{0}^{R}$ by $\tau$ and try to solve
(\ref{discrete_constraint}) along the lines of the treatment in
\cite{Bal_Gov_Mol_Paulo} of time-dependent Hamiltonians for
$\mathcal{A}_{\theta}(\mathbb{R}^{2})$. But for that we
need to know the spectrum $\mbox{spec}\,\hat{x}_{0}^{R}$ of $\hat{x}_{0}^{R}$
in $\mathcal{A}_{\theta}\left(\mathbb{R}\times S^{1},e^{i\frac{\varphi}{2\pi}},
e^{i\frac{\psi}{2\pi}}\right)$, since $\hat{H}$ is defined in $\tau$
only for $\tau\in \mbox{spec}\,\hat{x}_{0}^{R}$.

In $\mathcal{A}_{\theta}\left(\mathbb{R}\times
S^{1},e^{i\frac{\varphi}{2\pi}},
e^{i\frac{\psi}{2\pi}}\right)$, we
choose the following domain $D_{\varphi}\left(\hat{x}_{0}^{R}\right)$
for $\hat{x}_{0}^{R}$:
\begin{equation}\label{domain}
D_{\varphi}\left(\hat{x}_{0}^{R}\right) =
\left\{\hat{\alpha}\in\mathcal{A}_{\theta}\left(\mathbb{R} \times 
S^{1},e^{i\frac{\varphi}{2\pi}},e^{i\frac{\psi}{2\pi}}\right):\tilde{\alpha}_{n}\left(\omega+\frac{2\pi}{\theta}\right)=
e^{-i\varphi}\tilde{\alpha}_{n}\left(\omega\right)\right\} \,.
\end{equation}
For $\hat{\alpha}\in D_{\varphi}\left(\hat{x}_{0}^{R}\right)$,

$$
\hat{x}_{0}^{R}\hat{\alpha}  =  \hat{\alpha}\hat{x}_{0} =
e^{i\frac{\psi}{2\pi}\hat{x}_{1}}\sum_{n\in\mathbb{Z}}\int_{-\frac{\pi}{\theta}}^{+\frac{\pi}{\theta}}d\omega\,
\tilde{\alpha}_{n}(\omega)e^{in\hat{x}_{1}}
\left(-i\frac{\partial}{\partial\omega}e^{i\omega\hat{x}_{0}}\right)
$$

\begin{equation}
=-ie^{i\frac{\psi}{2\pi}\hat{x}_{1}}\sum_{n\in\mathbb{Z}}e^{in\hat{x}_{1}}\left[\tilde{\alpha}_{n}(\omega)
e^{i\omega\hat{x}_{0}}\right]_{-\frac{\pi}{\theta}}^{\frac{\pi}{\theta}}+
e^{i\frac{\psi}{2\pi}\hat{x}_{1}}
\sum_{n\in\mathbb{Z}}\int_{-\frac{\pi}{\theta}}^{+\frac{\pi}{\theta}}d\omega
\left(i\frac{\partial}{\partial\omega}\tilde{\alpha}_{n}(\omega)\right)
e^{in\hat{x}_{1}}e^{i\omega\hat{x}_{0}} \,.
\end{equation}
The first (surface) terms vanish by (\ref{domain}) and 
\begin{equation}
\hat{x}_{0}^{R}\hat{\alpha} =
e^{i\frac{\psi}{2\pi}\hat{x}_{1}}\sum_{n\in\mathbb{Z}}\int_{-\frac{\pi}{\theta}}^{+\frac{\pi}{\theta}}d\omega\,
\left(i\frac{\partial}{\partial\omega}\tilde{\alpha}_{n}(\omega)\right)
e^{in\hat{x}_{1}}e^{i\omega\hat{x}_{0}} \,.
\end{equation}
Hence if 

\begin{equation}
\hat{\alpha}
=e^{i\frac{\psi}{2\pi}\hat{x}_{1}}\int_{-\frac{\pi}{\theta}}^{+\frac{\pi}{\theta}}d\omega\,
e^{-i\omega\theta\left(n+\frac{\varphi}{2\pi}\right)}
e^{in\hat{x}_{1}}e^{i\omega\hat{x}_{0}} \,,
\end{equation}
then $\hat{\alpha}$ is an eigenvector of $\hat{x}_{0}^{R}$:
\begin{equation}\label{eigenvec}
\hat{x}_{0}^{R}\hat{\alpha}=\theta\left(n+\frac{\varphi}{2\pi}\right)\hat{\alpha}
\,,
\end{equation}
and for the spectrum of $\hat{x}_{0}^{R}$, we get
\begin{equation}
\textrm{spec} \, \hat{x}_{0}^{R} =
\theta\mathbb{Z}+\frac{\theta\varphi}{2\pi}\,.
\end{equation}
It is spaced in units of $\theta$.

We can now solve (\ref{discrete_constraint}):

\begin{equation}\label{solution_constraint}
\hat{\psi} = U\left(\hat{x}_{0}^{R},-\infty\right)
e^{i\frac{\psi}{2\pi}\hat{x}_{1}}\hat{\chi}\left(e^{i\hat{x}_{1}}\right)\,,
\end{equation}
where

$$U\left(\hat{x}_{0}^{R},-\infty\right)= 
U\left(\hat{x}_{0}^{R}-\theta\right) 
U\left(\hat{x}_{0}^{R}-2\theta\right)\dots\,,$$

\begin{equation}
U\left(\hat{x}_{0}^{R}-j\theta\right)=  
e^{-i\left(\hat{x}_{0}^{R}-j\theta\right)
\hat{H}\left(\hat{x}_{0}^{R}-j\theta\,,\,e^{i\hat{x}_{1}^{L}}\,,\,\hat{P}_{1}\right)}\,.
\end{equation}
Observables compatible with the Schr\"{o}dinger constraint can be
constructed as before.

\subsection*{\it 4. Remarks}

We point out that we can see the absence of
nontrivial null states in
$\tilde{\mathcal{H}}_{\theta}\left(e^{i\frac{\varphi}{2\pi}},
e^{i\frac{\psi}{2\pi}}\right)$ as before so that
the inner product becomes a true scalar product for
$\tilde{\mathcal{H}}_{\theta}\left(e^{i\frac{\varphi}{2\pi}},
e^{i\frac{\psi}{2\pi}}\right)$. Also, the Hilbert space
$\mathcal{H}_{\theta}\left(e^{i\frac{\varphi}{2\pi}},
e^{i\frac{\psi}{2\pi}}\right)$ obtained by completion of
$\tilde{\mathcal{H}}_{\theta}\left(e^{i\frac{\varphi}{2\pi}},
e^{i\frac{\psi}{2\pi}}\right)$ is independent of $m$ in
(\ref{same_inner_product}) while $\hat{x}_{0}^{L,R}$ do not act on
$\mathcal{H}_{\theta}\left(e^{i\frac{\varphi}{2\pi}},
e^{i\frac{\psi}{2\pi}}\right)$. 

Note that while $e^{-i\frac{2\pi}{\theta}\hat{x}_{0}^{R}}$ acts on 
$\mathcal{H}_{\theta}\left(e^{i\frac{\varphi}{2\pi}},
e^{i\frac{\psi}{2\pi}}\right)$, it is $e^{-i\varphi}\mathbb{I}$
because of (\ref{eigenvec}). So it cannot be the starting point to
define a time operator.

These remarks generalise to the other examples of discrete evolution
considered below.

\section*{IV. Noncommutative $\mathbb{R}^{3}$}

Here we show that the algebra $\hat{e}_{2}$ admits
a positive map. With that, one
can proceed to develop quantum physics.  

If $\hat{x}_{0}$, $\hat{x}_{a}$ ($a=1,2$) are time and space coordinate
functions in commutative spacetime, we call their noncommutative analogues
also by $\hat{x}_{0}$, $\hat{x}_{a}$. They fulfill the relations
\[
\left[\hat{x}_{a},\hat{x}_{b}\right] = 0
\,\,,\,\, a,b=1,2 \,,
\]
\begin{equation}\label{2+1_nc_relations}
\left[\hat{x}_{0},\hat{x}_{a}\right] =
i\theta\varepsilon_{ab}\hat{x}_{b}\,\,,\,\,\varepsilon_{12}=
-\varepsilon_{21}=1\,\,,\,\,\theta>0 \,.
\end{equation}
(\ref{2+1_nc_relations}) defines the Lie algebra of the two-dimensional
Euclidean group, and admits a $*$-operation:
$\hat{x}_{\mu}^{*}=\hat{x}_{\mu}$. Equally important, it admits the
time-translation automorphism
$U(\tau):U(\tau)\hat{x}_{0}=\hat{x}_{0}+\tau$. But it is not an inner
automorphism, $\hat{x}_{0}$ having no conjugate operator. 

Spatial translations are not automorphisms of (\ref{2+1_nc_relations}).
That means that momenta, free Hamiltonian or plane waves do not exist
for (\ref{2+1_nc_relations}).

The algebra $\hat{e}_{2}$
with relations (\ref{2+1_nc_relations}) has been treated in detail
by Chaichian et al. \cite{Chaichian}. As they observe, the operator 
\begin{equation}
\rho^{2} = \sum\hat{x}_{a}\hat{x}_{a}
\end{equation}
is in the center of $\hat{e}_{2}$.
We can fix its value to be $r^{2}$ in an IRR just as we fixed the
value of
$e^{-i\frac{2\pi}{\theta}\hat{x}_{0}} \in \mathcal{A}_{\theta}
\left(\mathbb{R} \times S^{1}\right)$.  
For $r^{2}>0$, we have the polar decomposition
\begin{equation}
\hat{x}_{1}\pm i\hat{x}_{2} = re^{\mp i\hat{x}} \,.
\end{equation}
Now
\begin{equation}
e^{i\hat{x}}\hat{x}_{0}=\hat{x}_{0}e^{i\hat{x}}+\theta
e^{i\hat{x}}\,,
\end{equation}
and $\hat{x}_{0}$, $e^{i\hat{x}}$ generate
$\mathcal{A}_{\theta}\left(\mathbb{R}\times S^{1}\right)$, the algebra
treated before. Hence we can borrow ideas from the treatment of
$\mathcal{A}_{\theta}\left(\mathbb{R}\times S^{1}\right)$. 

We briefly treat (\ref{2+1_nc_relations}) regarding $\hat{x}_{a}$
as generators of $C^{\infty}\left(\mathbb{R}^{2}\right)$ and
$\hat{x}_{0}/\theta$ as the generator of rotations in the $1-2$ plane. The
algebra will be realised by operators on the auxiliary Hilbert space
$L^{2}\left(\mathbb{R}^{2},d^{2}x\right)$ with its standard scalar
product $(\,.\,,\,.\,)$ where
\begin{equation}
\left(\alpha,\beta\right) = \int d^{2}x \, \alpha^{*}(x)\beta(x)\,.
\end{equation}
On this space, $\hat{x}_{a}$ acts by evaluation map,
\begin{equation}
\hat{x}_{a}\alpha(x) = x_{a}\alpha(x) \,,
\end{equation}
while $\hat{x}_{0}/\theta$ acts like angular momentum with 
\begin{equation}
e^{i2\pi \hat{x}_{0}/\theta} = \mathbb{I}\,.
\end{equation}
Then for the spectrum of $\hat{x}_{0}$,
\begin{equation}
\textrm{spec} \, \hat{x}_{0} = \theta\mathbb{Z} \,.
\label{spec_gravity}
\end{equation}

\noindent
Time is quantised in units of $\theta$ as for
$\mathcal{A}_{\theta}\left(\mathbb{R}\times S^{1}\right)$, but there
is no shift from $\theta\mathbb{Z}$ by a flux term
$\theta\varphi/2\pi$.  

There are also ray representations of the Euclidean group which are
representations of (\ref{2+1_nc_relations}), where the spectrum
$\theta\mathbb{Z}$ is shifted by a flux term
$\frac{\theta\varphi}{2\pi}$. Our discussion can be adapted to this
case as well.

We now give the positive map and inner product for
$\hat{e}_{2}$. 

The algebra
$\hat{e}_{2}$ is generated by
\begin{equation}
e^{i\omega\hat{x}_{0}} \,,\quad 
e^{i\vec{p}.\hat{x}}\,,\quad\vec{p}.\hat{x} =
p_{1}\hat{x}_{1}+p_{2}\hat{x}_{2} \,,\quad
\omega,p_{a}\in\mathbb{R} \,,
\end{equation}
where because of the spectral condition (\ref{spec_gravity}),
\begin{equation}
e^{i\left(\omega+\frac{2\pi}{\theta}\right)\hat{x}_{0}}=
e^{i\omega\hat{x}_{0}}\,.
\end{equation}
Thus we restrict $\omega$ according to 
\begin{equation}
| \omega|\le\frac{\pi}{\theta} \,.
\end{equation}

The general element of the algebra is 
\begin{equation}
\hat{\alpha} =\int d^{2}p \, 
\int_{-\frac{\pi}{\theta}}^{+\frac{\pi}{\theta}}d\omega\,
\tilde{\alpha}(\omega,\vec{p})e^{i\vec{p}.\hat{x}}
e^{i\omega\hat{x}_{0}} \,.
\end{equation}
The symbol we associate to $\hat{\alpha}$ is the function 
\[
\alpha:\theta\mathbb{Z}\times\mathbb{R}^{2}\rightarrow\mathbb{C}\,,
\]
\begin{equation}
\alpha(\theta n,x) = \int d^{2}p \,
\int_{-\frac{\pi}{\theta}}^{+\frac{\pi}{\theta}}d\omega\,
\tilde{\alpha}(\omega,\vec{p})\,e^{i\vec{p}.\vec{x}}\,
e^{i\omega\theta n}\,\,,\,\, n\in\mathbb{Z} \,.
\end{equation}
This gives the map
\begin{equation}
S_{\theta n}\left(\hat{\alpha}\right)=\int d^{2}x \,
\alpha(\theta n,x) \,.
\label{symbol_map}
\end{equation}
We can show that (\ref{symbol_map}) is a positive map. We have the
identity

\begin{equation}
e^{-i\omega\hat{x}_{0}}\hat{x}_{a}e^{i\omega\hat{x}_{0}}=
R_{ab}\left(\theta\omega\right)\hat{x}_{b} \,\,,\,\, 
R\left(\theta\omega\right) = 
\left(\begin{array}{cc}
\cos\left(\theta\omega\right) & \sin\left(\theta\omega\right)\\
-\sin\left(\theta\omega\right) & \cos\left(\theta\omega\right)
\end{array}\right) \,.
\end{equation}
A short calculation which uses this identity shows, in an obvious
manner, that 
\begin{equation}
S_{\theta n}\left(\hat{\alpha}^{*}\hat{\alpha}\right) =(2\pi)^{2}
\int d^{2}p\left|\int d\omega \,
\tilde{\alpha}\left(\omega,\vec{p}\right)
e^{i\omega\theta n}\right|^{2} \ge 0 \,.
\end{equation}
Thus an inner product for
$\hat{e}_{2}$ is
\begin{equation}
\left(\hat{\beta},\hat{\alpha}\right) =
S_{\theta n} \left(\hat{\beta}^{*}\hat{\alpha}\right) \,.
\end{equation}

\section*{V. The Noncommutative $\mathbb{R}\times S^{3}$}

The noncommutative $\mathbb{R}\times S^{3}\simeq\mathbb{R}\times SU(2)$ is denoted by
$\mathcal{A}_{\theta}\left(\mathbb{R}\times S^{3}\right)$. Section I
indicates its construction: we set the time operator $\hat{x}_{0}$
equal to $2\theta J_{3}^{R}$ where $\theta$ is the noncommutativity
parameter. With $\mathcal{C}^{\infty}\left(SU(2)\right)$
denoting the commutative algebra of functions on $SU(2)$, 
$\mathcal{A}_{\theta}\left(\mathbb{R}\times S^{3}\right)$ is generated
by $\mathcal{C}^{\infty}\left(SU(2)\right)$ and $\hat{x}_{0}$ with
relation (\ref{non_s3}).

Let $L^{2}\left(SU(2)\,,\,d\mu\right)$ denote the Hilbert space of
functions on $SU(2)$ with scalar product $(\cdot\,,\,\cdot)$ given by
the Haar measure $d\mu$:

\begin{equation}\label{haar}
(\hat{a}\,,\,\hat{b})=\int d\mu(s)\,\hat{a}^{*}(s)\,\hat{b}(s)\,.
\end{equation}
Then $\mathcal{A}_{\theta}\left(\mathbb{R}\times S^{3}\right)$ acts
naturally on this Hilbert space, $\mathcal{C}^{\infty}(SU(2))$ acting
by point-wise multiplication and $\hat{x}_{0}$ as the differential
operator $2\theta J_{3}^{R}$.

The spectrum $\mbox{spec}\,J_{3}^{R}$ of $J_{3}^{R}$ is $\mathbb{Z}/2$. Hence
$\mbox{spec}\,\hat{x}_{0}=\theta\mathbb{Z}$. Therefore

\begin{equation}
e^{i2\pi\hat{x}_{0}/\theta}=\mathbb{I}\,.
\end{equation} 

It follows that time evolution is quantised in units of $\theta$.

Furthermore

\begin{equation}
e^{i\left(\omega+\frac{2\pi}{\theta}\right)\hat{x}_{0}}=e^{i\omega\hat{x}_{0}}\,.
\end{equation}
Hence we can restrict $\omega$ to
$\left[-\frac{\pi}{\theta}\,,\frac{\pi}{\theta}\right]$ and
represent an element $\hat{\psi}$ of
$\mathcal{A}_{\theta}\left(\mathbb{R}\times S^{3}\right)$ as

\begin{equation}
\hat{\psi}=\int_{-\pi/\theta}^{\pi/\theta}d\omega\,\hat{\psi}_{\omega}\,e^{i\omega\hat{x}_{0}},
\quad \hat{\psi}_{\omega}\in \mathcal{C}^{\infty}(SU(2))\,.
\end{equation}

The symbol of $\hat{\psi}$ is the function
$\psi:(\mbox{spec}\,\hat{x}_{0}=\theta\mathbb{Z})\times SU(2)\longrightarrow\mathbb{C}$
defined by

\begin{equation}
\psi(\theta n\,,\,s)=\int_{-\frac{\pi}{\theta}}^{\frac{\pi}{\theta}}
d\omega\,\hat{\psi}_{\omega}(s)\,e^{i\omega\theta n}\,,\quad n\in \mathbb{Z}\,.
\end{equation}

The inner product can be obtained from an associated map $S_{\theta n}$:

\begin{equation}
S_{\theta n}(\hat{\psi})=\int d\mu(s)\,\psi(\theta n\,,s)\,.
\end{equation}
We can check using the right-invariance of the Haar measure that 

\begin{equation}
S_{\theta n}(\hat{\psi}^{*}\hat{\varphi})=
\int d\mu(s)\,\psi^{*}(\theta n,s)\,\varphi(\theta n,s)\,,
\end{equation}
where $\varphi$ is the symbol of $\hat{\varphi}$. Hence $S_{\theta n}$
is a positive map.

The rest of the treatment involving the Schr\"{o}dinger constraint
follows previous sections.

\section*{VI. On Energy Conservation}

We focus on time-independent Hamiltonians $\hat{H}$. In that case,
the Schr\"{o}dinger constraint such as (\ref{discrete_constraint}) is preserved by
$\hat{H}$,

\begin{equation}
\hat{\psi}\in \tilde{\mathcal{H}}_{\theta}\left(e^{i\frac{\varphi}{2\pi}}\,,\,e^{i\frac{\psi}{2\pi}}\right)
\Longrightarrow \hat{H}\hat{\psi}\in 
\tilde{\mathcal{H}}_{\theta}\left(e^{i\frac{\varphi}{2\pi}}\,,\,e^{i\frac{\psi}{2\pi}}\right)\,,
\end{equation}
and consequently $\hat{H}$ is an observable for
$\mathcal{A}_{\theta}\left(\mathbb{R}\times S^{1}\right)$. The same is
true for $\hat{e}_{2}$ and
$\mathcal{A}_{\theta}\left(\mathbb{R}\times S^{3}\right)$.

However time evolution involves

\begin{eqnarray}
U(\theta)=e^{-i\theta\hat{H}}\,,
\end{eqnarray}
its inverse and powers. It is the same for $\hat{H}$ and
$\hat{H}+\frac{2\pi}{\theta}$. Hence time evolution need conserve
energy only mod $\frac{2\pi}{\theta}$.

This energy nonconservation should show up in scattering and decay
processes. In either case, if $E_{i}$ and $E_{f}$ are initial and final
energies, then for $\theta=0$, energy conservation is enforced by the
factor 

\begin{equation}
\int_{-\infty}^{\infty}d\tau\,e^{-i\tau (E_{f}-E_{i})}=
2\pi\delta (E_{f}-E_{i})
\end{equation}
in the scattering matrix element. For quantised evolutions such as
ours, the factor becomes 

\begin{equation}\label{delta_circ}
\sum_{n\in\mathbb{Z}}e^{-in\theta
(E_{f}-E_{i})}=2\pi\delta_{S^{1}}\left[\theta (E_{f}-E_{i})\right]
\end{equation}
where $\delta_{S^{1}}$ is the $\delta$-function on $S^{1}$: 
$\delta_{S^{1}}(\theta +2\pi)=\delta_{S^{1}}(\theta)$. Thus from an
initial state of energy $E_{i}$, there can be transitions to energies
$E_{f}=E_{i}+\frac{2\pi}{\theta}n,\,\, n\in \mathbb{Z}$.

In specific models, the probability $P_{n}(E)$ for transitions from
$E_{i}=E$ to $E_{f}=E+\frac{2\pi}{\theta}n$ can be calculated. We
initiate the theory for this purpose in the next section. We are
looking for a manageable model for a specific calculation.

Suppose that we start with a state of sharp energy $E$ and let it
undergo multiple scattering. Let the probability for finding energy
$E+\frac{2\pi}{\theta}n$ after $k$ scatterings be
$P_{n}(E\,,\,k)$. Then

\begin{equation}\label{Markov}
P_{n}(E\,,\,k+1)=\sum_{m}P_{n-m}\left(E+\frac{2\pi}{\theta}m\,,\,1\right)\,P_{m}(E\,,\,k)
\end{equation}  
where

\begin{equation}
P_{n}(E\,,\,1)=P_{n}(E)\,.
\end{equation} 
Equation (\ref{Markov}) defines a Markov process with $P_{n}(E\,,\,1)$ giving
the rule for updating at each step. It is of considerable interest to
study $P_{n}(E\,,\,k)$ and its limit $k\rightarrow \infty$.

We remark that the limiting distribution $P_{n}(E\,,\,\infty)$ may be
of use to provide bounds on $\theta$ in conjunction with cosmological
data. Presumably distant star or quasar signals arrive at us after a
large number of scattering processes. We can imagine estimating their
frequency dispersion after accounting for energy loss by standard
$\theta=0$ effects, and getting information on $\theta$ therefrom.

\section*{VII. Scattering Theory}

We consider only a situation where the Hamiltonian $\hat{H}$ is
time-independent.

The transition amplitude from the in state vector $\lvert
+\,,\,\alpha\rangle$ with label $\alpha$ to an out state vector 
$\lvert -\,,\,\beta\rangle$ with label $\beta$ defines the matrix element
$\mathcal{S}_{\beta\alpha}$ of the $S$-matrix $\mathcal{S}$:

\begin{equation}
\mathcal{S}_{\beta\alpha}=\langle -\,,\,\beta\lvert +\,,\,\alpha\rangle\,.
\end{equation}

Let $\hat{H}_{0}$ be the ``free'' or ``comparison'' Hamiltonian. Then 
$\lvert +\,,\,\alpha\rangle$ has the property 

\begin{equation}\label{past}
U(\theta)^{N}\lvert
+\,,\,\alpha\rangle=U_{0}(\theta)^{N}\lvert\alpha\rangle
\quad \mbox{as}\,\, N\rightarrow-\infty, \quad \mbox{with}\,\, N\in\mathbb{Z}
\end{equation}
where 

\begin{equation}
U_{0}(\theta)=e^{-i\theta\hat{H}_{0}}\,.
\end{equation}
The meaning of (\ref{past}) is that in the distant past,
$\lvert +\,,\,\alpha\rangle$ evolves like the free evolution of the vector $\lvert
\alpha\rangle$. 

The label $\alpha$ can be given a meaning in terms of
observables of the free system such as energy.

The limit involved requires care. It is to be understood in the strong
sense. It defines the M\o ller operator

\begin{equation}
\Omega^{+}=\lim_{\buildrel{N \to -\infty\,,}\over{N\in\mathbb{Z}}} U(\theta)^{-N}U_{0}(\theta)^{N}
\end{equation}
with the properties

\begin{eqnarray}\label{properties}
\Omega^{+}\lvert\alpha\rangle=\lvert +\,,\,\alpha\rangle\,,
\end{eqnarray}

\begin{equation}\label{inter}
\Omega^{+}e^{-i\theta\hat{H}_{0}}=e^{-i\theta\hat{H}}\Omega^{+}\,.
\end{equation}

Equation (\ref{properties}) follows from (\ref{past}) while the proof
of (\ref{inter}) is as follows: 

\begin{equation}\label{proof}
\Omega^{+}e^{-i\theta\hat{H}_{0}}=\lim_{\buildrel{N\to -\infty\,,}\over{N\in\mathbb{Z}}}
U(\theta)^{-N}U_{0}(\theta)^{N+1}=
\lim_{\buildrel{N'\to -\infty\,,}\over{N'\in\mathbb{Z}}}
U(\theta)^{-(N'-1)}U_{0}(\theta)^{N'}=
e^{-i\theta\hat{H}}\Omega^{+}\,.
\end{equation}

Thus $\Omega^{+}$ intertwines the quantised evolutions due to
$\hat{H}_{0}$ and $\hat{H}$.

For $\theta=0$, time $t$ is continuous. In that case, (\ref{inter}) is
replaced by 

\begin{equation}
\Omega^{+}e^{-it\hat{H}_{0}}=e^{-it\hat{H}}\Omega^{+}\,.
\end{equation}
So for $\theta=0$, by differentiating in $t$, we get the stronger
result 

\begin{equation}
\Omega^{+}\hat{H}_{0}=\hat{H}\Omega^{+}\,.
\end{equation}
But we cannot get such a stronger equation from (\ref{inter}) for $\theta\neq
0$. This is yet another indication that for $\theta\neq 0$, energy is
conserved only mod $\frac{2\pi}{\theta}$.

Just as $\lvert +\,,\,\alpha\rangle$ fulfills the Schr\"{o}dinger
constraint involving $\hat{H}$, $\lvert\alpha\rangle$ fulfills the
Schr\"{o}dinger constraint involving $\hat{H}_{0}$ as follows from
(\ref{inter}):

\begin{equation}
e^{-i\theta\hat{P}_{0}}\lvert\alpha\rangle=
e^{-i\theta\hat{H}_{0}}\lvert\alpha\rangle\,.
\end{equation}
So scalar products involving $\lvert\alpha\rangle$'s are also
time-independent and admit a general solution of a form such as (\ref{solut}).

In a similar way, if

\begin{equation}
\Omega^{-}=\lim_{\buildrel {M \to \infty\,,} \over {M\in\mathbb{Z}}} U(\theta)^{-M}U_{0}(\theta)^{M}\,,
\end{equation}
then

\begin{equation}
\Omega^{-}\lvert\beta\rangle=\lvert -\,,\,\beta\rangle\,,
\end{equation}

\begin{equation}
\Omega^{-}e^{-i\theta\hat{H}_{0}}=
e^{-i\theta\hat{H}}\Omega^{-}\,.
\end{equation}
Hence

\begin{equation}
\mathcal{S}_{\beta\alpha}=\lim_{\buildrel{M\to\infty\,,}\over{\buildrel{N\to -\infty\,,}\over{M,N\in\mathbb{Z}}}}
\langle\beta\lvert U_{0}(\theta)^{-M}U(\theta)^{M-N}U_{0}(\theta)^{N}\rvert\alpha\rangle
:=\lim_{\buildrel{M\to\infty\,,}\over{\buildrel{N\to -\infty\,,}\over{M,N\in\mathbb{Z}}}}
\langle\beta\lvert U_{I}(\theta,M,N)\rvert\alpha\rangle\,,
\end{equation}

\begin{equation}\label{M_N}
U_{I}(\theta,M,N) = U_{0}(\theta)^{-M}U(\theta)^{M-N}U_{0}(\theta)^{N}
= e^{iM\theta\hat{H}_{0}}e^{-i(M-N)\theta\hat{H}}e^{-iN\theta\hat{H}_{0}}\,.
\end{equation}

In commutative physics, where $\theta=0$, the corresponding expression
$U_{I}(t\,,\,t')$ is

\begin{equation}\label{commutative}
U_{I}(t\,,t') = e^{it\hat{H}_{0}}e^{-i(t-t')\hat{H}}e^{-it'\hat{H}_{0}}
=T\,exp\left\{-i\int_{t'}^{t} d\tau\,\hat{H}_{I}(\tau)\right\}\,,
\end{equation}

\begin{equation}\label{H_I}
\hat{H}_{I}(\tau)=e^{i\hat{H}_{0}\tau}(\hat{H}-\hat{H}_{0})e^{-i\hat{H}_{0}\tau}\,,
\end{equation}
$T$ denoting time-ordering, the interaction representation $S$-matrix
being $U_{I}(\infty\,,\,-\infty)$.

Comparison of (\ref{M_N}) and (\ref{commutative}) shows that 

\begin{equation}\label{comp}
U_{I}(\theta,M,N)=T\,exp\left\{-i\int_{N\theta}^{M\theta}d\tau
\,\hat{H}_{I}(\tau)\right\}\,\,,
\end{equation}

\begin{equation}
\hat{H}_{I}(\tau)=e^{i\hat{H}_{0}\tau}(\hat{H}-\hat{H}_{0})e^{-i\hat{H}_{0}\tau}.
\end{equation}

For $\theta=0$, (\ref{commutative}) has a power series expansion in
$\hat{H}_{I}$. But there is a problem with such an expansion of
(\ref{M_N}): $U(\theta)$, $U_{0}(\theta)$ and $U_{I}(\theta,M,N)$
are invariant under {\it separate} shifts of $\hat{H}$ and
$\hat{H}_{0}$ by $\pm \frac{2\pi}{\theta}$, however
$\hat{H}_{I}(\tau)$ and hence the terms of the perturbation series are
invariant only under the {\it joint} shift of both by the same
amount, the joint shift leaving $\hat{H}_{I}(\tau)$ invariant. 
Thus perturbative approximation disturbs an essential feature 
of quantised evolution.

It remains to find a substitute for perturbation theory. Perhaps an 
approximation based on the $K$-matrix formalism and effective range 
expansion \cite{matrix}, \cite{range} may be acceptable.

\section*{VIII. On Quantum Fields}

As the spacetime algebras of our interest admit only quantised time
evolutions as automorphisms, a field cannot be the solution of a
Klein-Gordon or Dirac equation. We need another approach to quantising
spacetime fields for purposes of constructing quantum fields.

One way is to define the quantum field $\hat{\Phi}$ by expanding it in
a basis of orthonormal solutions of the Schr\"{o}dinger
constraint. The coefficients of the expansion would be
annihilation operators. This is a common approach in
condensed matter theory.

For specificity consider
$\mathcal{A}_{\theta}\left(\mathbb{R}\times S^{1}\right)$ and the ``free''
Hamiltonian

\begin{equation}
\hat{H}_{0}=\frac{\hat{P}_{1}^{2}}{2M}\,.
\end{equation}
In that case,
$\tilde{\mathcal{H}}_{\theta}\left(e^{i\frac{\varphi}{2\pi}}\,,\,e^{i\frac{\psi}{2\pi}}\right)$
of (\ref{discrete_constraint}) is spanned by

\begin{equation}
\hat{\psi}_{n}=\frac{1}{\sqrt{2\pi}}e^{i\left(n+\frac{\psi}{2\pi}\right)\hat{x}_{1}}e^{-i\omega_{n}\hat{x}_{0}}\,,
\end{equation}
where

\begin{eqnarray}
\omega_{n}=\frac{1}{2M}\left(n+\frac{\psi}{2\pi}\right)^{2}\,,
\end{eqnarray}

\begin{eqnarray}
\hat{H}_{0}\hat{\psi}_{n}=\frac{1}{2M}\left(n+\frac{\psi}{2\pi}\right)^{2}\hat{\psi}_{n}\,,
\end{eqnarray}

\begin{equation}
(\hat{\psi}_{m}\,,\,\hat{\psi}_{n})_{\theta\left(m+\frac{\varphi}{2\pi}\right)}=
\delta_{mn}\,.
\end{equation}

We can now write

\begin{equation}
\hat{\Phi}=\sum_{n}a_{n}\hat{\psi}_{n}, \quad\left[a_{n}\,,\,a_{m}^{\dagger}\right]=\delta_{nm}
\end{equation}
where $\hat{\Phi}$ describes a free ``nonrelativistic'' spin-zero field. (We
will not consider higher spins in this sketch.)

The second-quantised free Hamiltonian associated with $\hat{\Phi}$ is

\begin{equation}
\hat{\hat{H}}_{0}=\sum_{n}\omega_{n}a^{\dagger}_{n}a_{n}\,.
\end{equation}
$\hat{\Phi}$ fulfills the second-quantised Schr\"{o}dinger constraint:

\begin{equation}
e^{-i\theta\hat{P}_{0}}\hat{\Phi}=U_{0}(\theta)^{-1}\hat{\Phi}U_{0}(\theta)\,,
\end{equation}

\begin{equation}
U_{0}(\theta)\equiv e^{-i\theta\hat{\hat{H}}_{0}}\,.
\end{equation}

The next step is to introduce an interaction Hamiltonian. We follow
earlier works \cite{Doplicher}, \cite{Bal_Gov_Mol_Paulo} in this
regard. An example of an interaction Hamiltonian in interaction
representation is 

\begin{equation}
\hat{\hat{H}}_{I}(\tau)=\,:\,e^{i\tau\hat{\hat{H}}_{0}}\lambda
S_{\theta\left(m+\frac{\varphi}{2\pi}\right)}
\left(\hat{\Phi}^{\dagger}\hat{\Phi}\,\hat{\Phi}^{\dagger}\hat{\Phi}\right)
e^{-i\tau\hat{\hat{H}}_{0}}\,:
\end{equation}
where $:\,\cdot\,:$ denotes normal ordering of $a_{n}$,
$a_{n}^{\dagger}$.

$U_{I}(\theta,M,N)$ follows from (\ref{comp}):

\begin{equation}\label{problem}
U_{I}(\theta,M,N)=T\,exp\left\{-i\int_{N\theta}^{M\theta}
d\tau\,\hat{\hat{H}}_{I}(\tau)\right\}\,\,,\,\,M, N\in\mathbb{Z}\,,
\end{equation}
the $S$-matrix being 

\begin{equation}
\mathcal{S}=\lim_{\buildrel{M \to \infty\,,}\over{\buildrel{N \to
-\infty\,,}\over{M,N\in\mathbb{Z}}}}\, 
U_{I}(\theta,M,N)\,.
\end{equation}
As before, perturbation
series, term by term, is not invariant under the shifts of 
$\hat{\hat{H}}_{I}(\tau)$ by $\pm\frac{2\pi}{\theta}$, whereas
(\ref{problem}) is. That leaves us with a problem. 

It is also important to know if and how $\mathcal{S}$ depends on
$\left(m+\frac{\varphi}{2\pi}\right)$.

\begin{acknowledgments}
This work was  supported by DOE under contract number DE-FG02-85ER40231,
by NSF under contract number INT9908763 and by FAPESP, Brazil. 
\end{acknowledgments}



\begin{thebibliography}{10}

\bibitem{Doplicher}S. Doplicher, K. Fredenhagen and J. Roberts, 
{\it Space-Time Quantization Induced by Classical Gravity},
Phys. Lett. B \textbf{331}, 39-44 (1994) ;
S. Doplicher, K. Fredenhagen and J. Roberts, {\it The Quantum
Structure of Spacetime at the Planck Scale and Quantum Fields}, 
Comm. Math. Phys. \textbf{172}, 187 (1995) [hep-th/0303037].
%
\bibitem{Chaichian}M. Chaichian, A. Demichev, P. Presnajder and
A. Tureanu, {\it  Space-Time Noncommutativity, Discreteness of Time and Unitarity},
Eur. Phys. J. C \textbf{20,} 767-772 (2001) [hep-th/0007156]. 
%
\bibitem{Matschull}H.-J. Matschull and M. Welling, {\it Quantum Mechanics
of a Point Particle in 2+1 Dimensional Gravity}, Class. Quant. Grav. {\bf
15}, 2981-3030 (1998) [gr-qc/9708054].
%
\bibitem{Hooft}G. 't Hooft, {\it Canonical Quantization of Gravitating
Point Particles in 2+1 Dimensions}, Class. Quant. Grav. {\bf 10}, 1653-1664
(1993) [gr-qc/9305008] ;
G. 't Hooft, {\it Quantization of Point Particles in 2+1 Dimensional
Gravity and Space-Time Discreteness}, Class. Quant. Grav. \textbf{13}, 1023-1040 
(1996) [gr-qc/9601014].
%
\bibitem{Welling} M. Welling, {\it One-Particle Hilbertspace of 2+1
Dimensional Gravity Using Non-Commuting Coordinates},
Nucl. Phys. Proc. Suppl. \textbf{57}, 346-349
(1997) [gr-qc/9703057] ;
M. Welling, {\it Two Particle Quantummechanics in 2+1 Gravity Using
Non Commuting Coordinates},
Class. Quant. Grav. \textbf{14}, 3313-3326 (1997) [gr-qc/9703058].
%
\bibitem{J_Matschull}H.-J. Matschull, {\it The Phase Space Structure
of Multi Particle Models in 2+1 Gravity},
Class. Quant. Grav. {\bf 18}, 3497-3560 (2001) [gr-qc/0103084].
%
\bibitem{Bal_Gov_Mol_Paulo}A. P. Balachandran, T. R. Govindarajan,
C. Molina and P. Teotonio-Sobrinho, {\it Unitary Quantum Physics with
Time-Space Noncommutativity}, [hep-th/0406125].
%
\bibitem{Bal_Gupta} A.P. Balachandran, K.S. Gupta and S. Kurkcuoglu, {\it
in preparation}.
%
\bibitem{Bal_Chandar} A. P. Balachandran and L. Chandar, {\it Discrete
Time from Quantum Physics}, 
Nucl. Phys. B \textbf{428}, 435-448 (1994) [hep-th/9404193].
%
\bibitem{Rieffel} M.A. Rieffel, {\it Deformation Quantization for
Actions of $\mathbb{R}^d$}, American Mathematical Society, Providence (1993).
%
\bibitem{Connes_Landi} A. Connes and G. Landi, {\it Noncommutative
Manifolds the Instanton Algebra and Isospectral Deformations}, 
Comm. Math. Phys. {\bf 221}, 141-159 (2001) [math.QA/0011194]. 
%
\bibitem{matrix} R.G. Newton, {\it Scattering Theory of Waves and Particles},
McGraw-Hill Book Company, New York (1966).
%
\bibitem{range} M.L. Goldberger and K.M. Watson, {\it Collision
Theory}, Wiley, New York (1964).

\end{thebibliography}
\end{document}